**Weaving Life into Regolith: Engineered Autotrophic–Heterotrophic Consortia for Autonomous Biofabrication from Granular Feedstocks**

Nisha Rokaya [a], Erin C. Carr [b], Kumar Shrestha [c], Richard A. Wilson [a], Yong Huang [d], Congrui Jin [e*]

[a] Department of Plant Pathology, University of Nebraska–Lincoln, Lincoln, NE 68583

[b] School of Biological Sciences, University of Nebraska–Lincoln, Lincoln, NE 68588

[c] Agricen Science, Pilot Point, TX 76258

[d] Department of Mechanical and Aerospace Engineering, University of Florida, Gainesville, FL 32611

[e] Department of Engineering Technology and Industrial Distribution, Texas A&M University, College Station, TX 77843

[*] Corresponding author: Congrui Jin; E-mail addresses: jincongrui@tamu.edu

**Abstract**

Long-duration human missions to Mars will require autonomous systems capable of converting in situ resources into structural materials, tools, and functional components. More broadly, such systems represent a class of resource-limited bioprocesses relevant to extreme-environment manufacturing. Here, we investigate engineered autotrophic–heterotrophic consortia, inspired by lichen biology, as a platform for autonomous biofabrication from granular feedstocks. We experimentally screened filamentous fungi and paired them with diazotrophic cyanobacteria to identify mutually supportive consortia capable of sustained growth and biomineral production in the presence of Martian regolith simulant as the primary inorganic substrate, without external organic carbon or nitrogen inputs. Selected co-cultures exhibited evidence of metabolic coupling, and untargeted metabolomic analysis revealed coordinated reprogramming consistent with integrated carbon and nitrogen metabolism within the consortia. These systems facilitated mineral consolidation of regolith particles, demonstrating the feasibility of near–closed-loop biomineral production under resource-limited conditions. While integration with additive manufacturing remains conceptual, this study establishes a framework for engineering self-sustaining microbial consortia for biomaterials production and highlights opportunities for coupling metabolism with material synthesis in both extraterrestrial and terrestrial environments.

## 1. Introduction

The development of autonomous bioprocesses capable of operating under resource-limited conditions represents a central challenge in biotechnology and bioengineering. Designing biological systems that can sustain growth, maintain functionality, and produce materials using minimal inputs remains an open problem in bioprocess engineering. Mars provides an extreme testbed for such systems.

The Martian environment is characterized by a $CO_2$-dominated low-pressure atmosphere, large temperature fluctuations, and persistent ionizing radiation, i.e., conditions traditionally considered prohibitive to complex terrestrial life. However, recent simulation studies challenge this assumption. Lichens, i.e., stable associations between fungal mycobionts and photosynthetic algae or cyanobacteria, have demonstrated metabolic resilience under controlled Martian simulation environments. In particular, metabolically active lichens exposed to simulated atmospheric composition, pressure regimes, temperature cycling, and radiation doses approximating surface conditions retained fungal metabolic activity throughout the experiment (Skubała et al. 2025). These findings suggest that cooperative biological systems may endure environmental extremes beyond the tolerance limits of isolated organisms.

Beyond survival, lichens provide a compelling model for biologically mediated mineral transformation. Through biochemical weathering, secretion of organic acids, ion chelation, and biomineral precipitation, lichens actively interact with and modify mineral substrates (Banfield et al. 1999; Chen et al. 2000; Bjelland et al. 2002). These processes reflect a broader capacity for biologically driven mineral reorganization and stabilization, arising from metabolic complementarity: photoautotrophic carbon fixation supports heterotrophic growth, while fungal architectures stabilize microenvironments and facilitate mineral interactions. This cooperative organization suggests a framework for engineering microbial consortia in which metabolism and mineral transformation are intrinsically coupled.

Long-duration human missions to Mars require autonomous manufacturing systems capable of converting local resources into structural materials, tools, and habitat components. Transporting large quantities of construction materials from Earth is energetically prohibitive, making regolith, the ubiquitous granular material covering the Martian surface, a practical feedstock (Karacasulu et al. 2025). Current in situ resource utilization (ISRU) strategies emphasize physicochemical approaches such as regolith sintering, melting, or chemical extraction, which demand substantial energy inputs and complex hardware. Biological alternatives, including microbially induced calcite precipitation (MICP), have demonstrated the feasibility of consolidating granular substrates under terrestrial conditions (Dikshit et al. 2022). However, such systems rely on exogenous nutrients, externally supplied reactants, or metabolically fragile monocultures, limiting their applicability in resource-constrained environments.

A key unmet challenge is the development of closed autotrophic–heterotrophic consortia capable of sustained growth and mineral consolidation without external organic inputs. Inspired by lichen biology, we hypothesize that engineered photoautotroph–heterotroph consortia can achieve this functionality. In such systems (Figure 1), diazotrophic cyanobacteria fix atmospheric carbon and nitrogen, generating organic carbon, reduced nitrogen species, and oxygen, while increasing carbonate availability through photosynthetic activity. Filamentous fungi, in turn, bind divalent metal ions such as $Ca^{2+}$ to their cell walls, serving as nucleation sites for biomineral precipitation such as $CaCO_3$ and contributing to structural reinforcement. These complementary metabolic roles couple nutrient cycling with mineral formation, defining a self-sustaining biological process for regolith consolidation under minimal-input conditions.

Leveraging recent advances in cell-laden bioinks for 3D printing (Shahriar et al., 2025; Ning et al., 2026; Habib et al., 2026), direct ink writing offers a potential route for single-step fabrication. Bioinks can be formulated from cyanobacterial cells, fungal cells, and regolith particles dispersed in an inorganic growth medium, with rheological properties tuned using biopolymers such as alginate (Dubbin et al., 2016). Cyanobacteria naturally produce alginate as part of their extracellular polymeric substances (EPS), and under certain conditions it can constitute a substantial fraction of the matrix (Saad et al., 2023). Physical crosslinking of alginate enables initial particle binding and shape retention during printing, while subsequent microbially mediated biomineralization provides permanent consolidation of the printed structure. In this framework, additive manufacturing and microbial metabolism operate as coupled processes: printing defines geometry, while microbial activity drives material strengthening over time. This integration establishes a pathway toward autonomous biofabrication systems in which structure formation and material consolidation emerge from a unified, biologically driven process.

**2. Motivation and Innovation**

During the past decades, MICP has emerged as a promising biological strategy for consolidating granular materials. One prominent example is the work of bioMASON (Ednie-Brown, 2013), which employed ureolytic bacteria to cement sand particles into masonry units through repeated flushing with $Ca^{2+}$ and urea solutions. In this approach, bacterial urease activity hydrolyzes urea to increase carbonate concentration, resulting in $CaCO_3$ precipitation that binds particles together. While effective under terrestrial manufacturing conditions, this approach depends on continuous nutrient inputs, chemical replenishment, and tightly controlled processing steps.

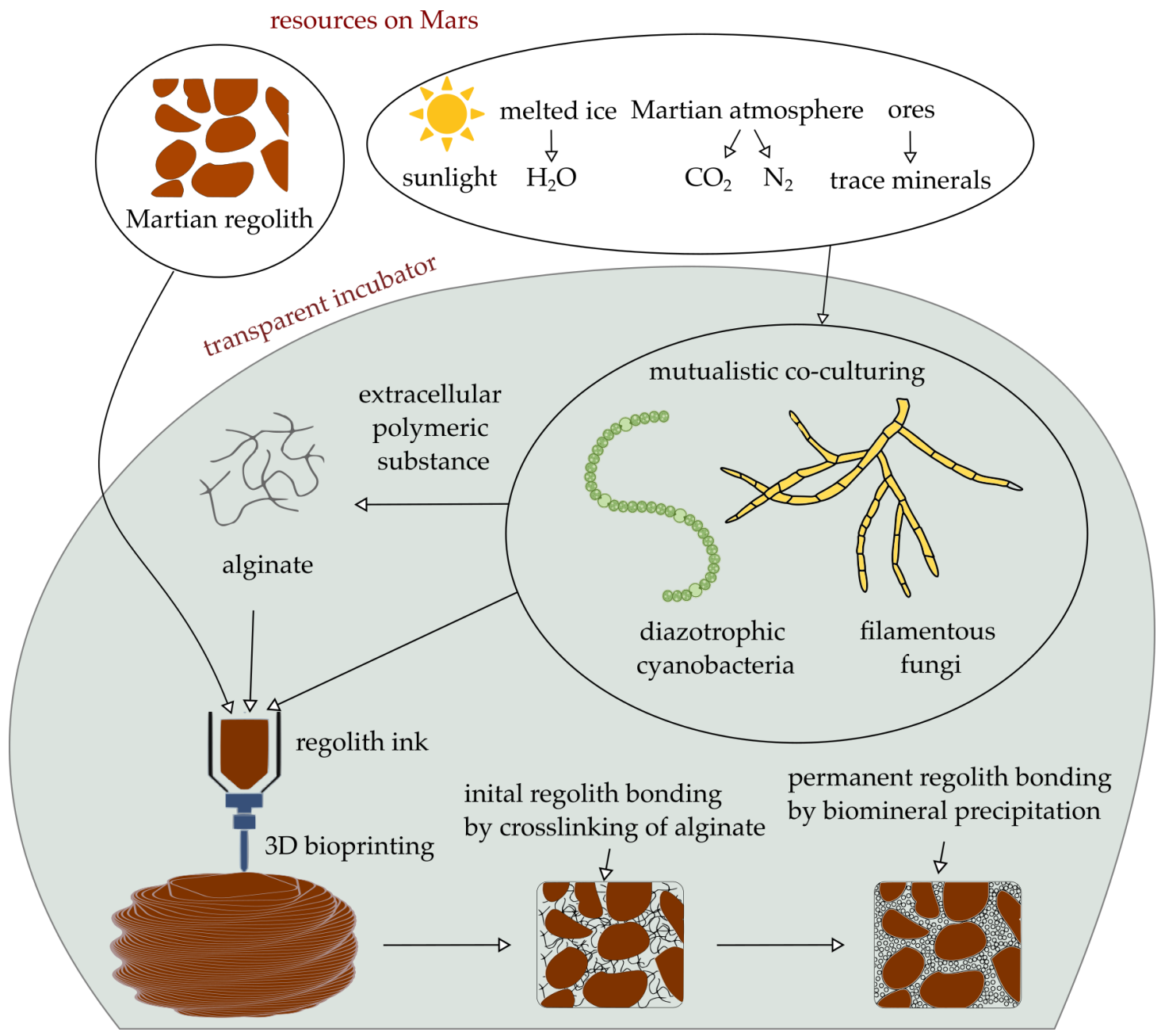

**FIGURE 1.** Schematic illustration of the proposed biological approach to Martian biomanufacturing.

To fabricate structures with complex geometries, powder bed–based 3D printing has been coupled with microbially induced calcite precipitation (MICP) to enable spatially patterned deposition of sand and urease-active $CaCO_3$ powder containing bacteria (Nething et al., 2020) (Figure 2A). After printing, structures are exposed to $Ca^{2+}$ and urea solutions to induce mineralization in targeted regions, while unbound sand is removed. Similarly, ureolytic bacteria have been incorporated into pre-printed polymeric lattices, where $CaCO_3$ precipitation occurs within inert mesh architectures (Figure 2B), but such implementations typically require post-processing steps, such as high-temperature annealing, to achieve mechanically robust structures (Xin et al., 2021). Although these systems demonstrate geometric control and biological mineral formation, they remain inherently multi-step processes that increase operational complexity, limit scalability, and hinder deployment in resource-limited or extreme environments.

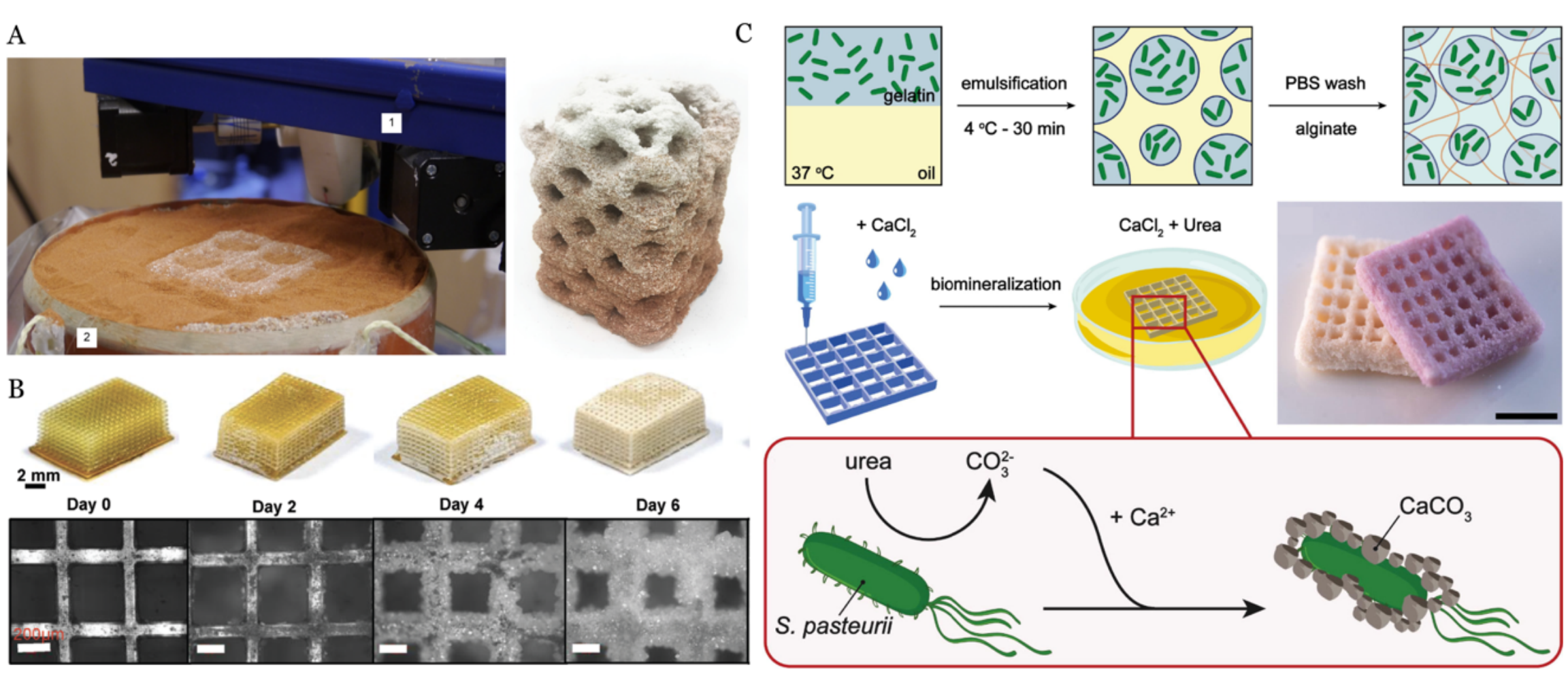

**FIGURE 2.** (A) Spatially patterned structures fabricated by using powder bed-based 3D printing

(Nething et al., 2020). 1: The print head. 2: The print bed. (B) Biominerals precipitated by ureolytic bacteria into 3D printed meshes (Xin et al., 2021). (C) Structures printed by microbe-loaded bioink in a single-step process via direct ink writing (Hirsch et al., 2023). Scale bar is 10 mm.

More recently, advances in cell-laden bioinks have enabled single-step extrusion-based fabrication. Hirsch et al. (2023) developed microbe-loaded bioinks capable of direct ink writing (Figure 2C), in which microbial cells and granular substrates are formulated into rheologically tunable inks and extruded through a moving nozzle. In such systems, microbial mineralization progressively converts a pliable scaffold into a load-bearing composite. Nevertheless, in such ureolysis-based implementations, continued mineral growth typically requires immersion in biomineralizing solutions containing $Ca^{2+}$ and urea, thereby maintaining dependence on externally supplied reactants. These constraints underscore the need for biologically autonomous and metabolically self-sustaining mineralization strategies compatible with extraterrestrial resource limitations. For Martian applications, a viable system must operate using locally available inputs, including sunlight, $CO_2$, $N_2$, water, and regolith-derived ions, while minimizing external supplementation and process complexity.

To address these limitations, this study advances an alternative strategy based on engineered autotrophic–heterotrophic consortia capable of sustained growth and biomineral production under minimal-input conditions (Figure 3). Because Martian regolith simulants are moderately to strongly alkaline, the initial phase involved screening alkaliphilic strains for survivability in regolith-containing media. Strains exhibiting robust growth were subsequently evaluated in co-culture interaction assays under progressively nutrient-limited conditions (C+N+, C+N−, C−N+, and C−N−) to assess metabolic complementarity. Monocultures and co-cultures were cultivated in inorganic liquid media supplemented with regolith simulant, air, and light, with or without added organic carbon or nitrogen sources. Microbial activity was quantified using resazurin reduction assays, phycocyanin fluorescence measurements, and fungal plating on selective media. Co-culture systems demonstrating sustained activity under carbon- and nitrogen-limited (C−N−) conditions were identified as candidates for closed-loop biomineral production and subsequently selected for untargeted metabolomic analysis to elucidate metabolic reprogramming and nutrient exchange dynamics. Through this staged framework encompassing survivability screening, co-culture interaction assessment, and metabolomic characterization, the study establishes a biological foundation for autonomous, lichen-inspired mineral consolidation in regolith-based environments.

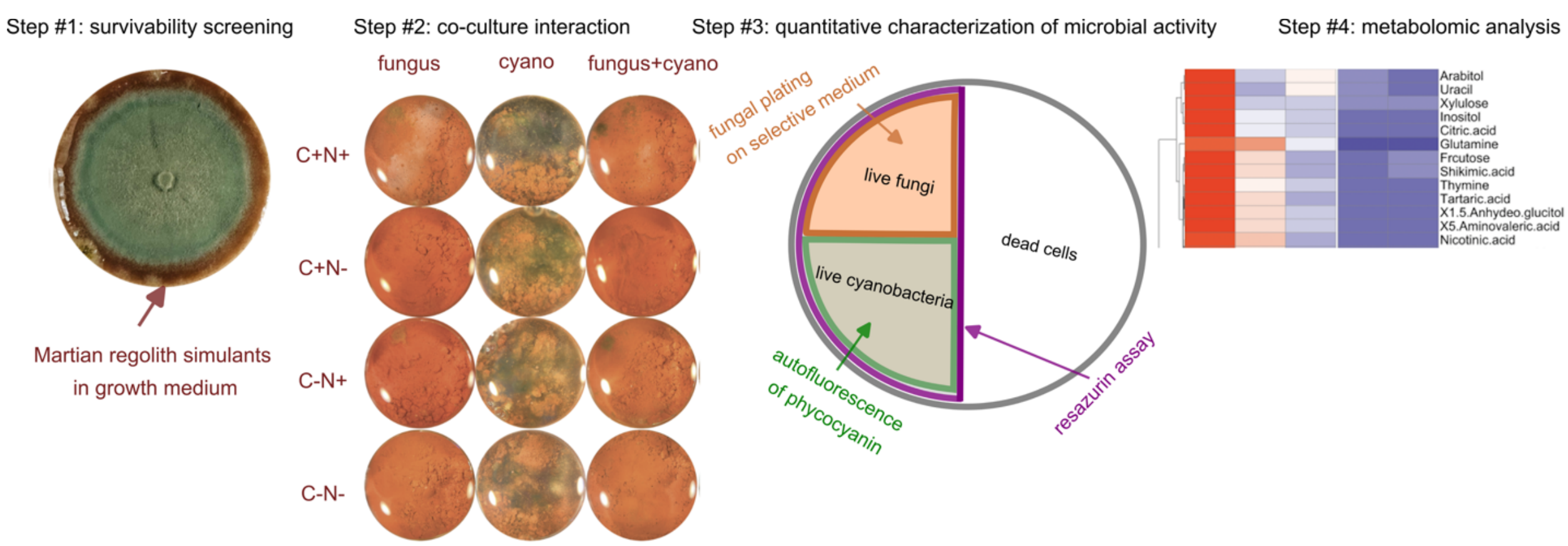

**FIGURE 3.** Experimental workflow for the development of regolith-compatible autotrophic–heterotrophic consortia.

## 3. Results and Discussion

### 3.1. Survivability Screening

To identify fungal partners capable of tolerating regolith-relevant geochemical conditions, fourteen filamentous fungal strains obtained from international culture collections (ATCC, CBS, and FGSC) were selected for growth screening on potato dextrose agar (PDA) supplemented with sterilized Enhanced Mojave Mars Simulant (MMS-2). Colony expansion was monitored over a 14-day incubation period (Figure S1), and radial growth rates and terminal pH values were recorded (Table 1).

The majority of strains demonstrated substantial tolerance to MMS-2 exposure. Ten of the fourteen fungi exhibited robust growth in the presence of regolith simulant and were therefore advanced to subsequent mutualism testing. Microscopic examination revealed abundant conidiation and well-formed conidiophores comparable to those observed on control plates lacking simulant, indicating that regolith exposure did not disrupt reproductive morphology. The highest growth rate (6.03 mm/day) was observed for *T. viride* (HF3MP), followed closely by *T. reesei* and *T. viride* (HF3TV). Four strains showed markedly reduced growth under MMS-2 conditions despite performing well on PDA alone, suggesting differential tolerance to alkaline or mineral-associated stressors. Sterile agar plug controls confirmed the absence of contamination or spontaneous growth.

The initial pH of the growth medium was 8.8, reflecting the alkaline character of the regolith simulant. Following incubation, all inoculated plates exhibited some degree of acidification, consistent with fungal metabolic activity and organic acid secretion. The lowest final pH (5.9) was recorded for *A. niger*, which also exhibited strong radial growth, suggesting active metabolic modification of the surrounding substrate. In contrast, *P. inflatum*, which showed no measurable growth, maintained a pH near the initial value (8.7), reinforcing the relationship between metabolic activity and medium acidification.

These results demonstrate that a substantial fraction of selected filamentous fungi can survive and remain metabolically active in alkaline regolith-simulant environments. Importantly, the observed pH shifts suggest active biochemical interaction with mineral substrates, a trait relevant to biomineralization and substrate transformation. The identified tolerant strains thus constitute viable heterotrophic candidates for constructing regolith-compatible autotrophic–heterotrophic consortia.

**Table 1.** The growth rates and pH measurements after 14-day incubation.

| Tested Fungal Strain | Growth Rate (mm/day) | pH | Selected? |
|---|---|---|---|
| *Aspergillus crystallinus* (ATCC16833) | 3.09 | 7.8 | No |
| *Aspergillus restrictus* (ATCC16912) | 5.32 | 8.1 | Yes |
| *Penicillium atramentosum* (ATCC10104) | 4.07 | 8.4 | Yes |
| *Hemicarpenteles paradoxus* (ATCC16918) | 2.38 | 7.8 | No |
| *Penicillium stipitatus* (ATCC10500) | 3.67 | 6.8 | Yes |
| *Penicillium inflatum* (ATCC48994) | 0.00 | 8.7 | No |
| *Penicillium hirayamae* (ATCC18312) | 1.97 | 6.1 | No |
| *Aspergillus malodoratus* (ATCC16834) | 3.17 | 7.5 | Yes |
| *Penicillium brefeldianum* (CBS235.81) | 4.33 | 7.8 | Yes |
| *Aspergillus niger* (ATCC16888) | 6.00 | 5.9 | Yes |
| *Aspergillus flavus* (ATCC9643) | 5.03 | 7.7 | Yes |
| *Trichoderma reesei* (ATCC13631) | 6.01 | 8.1 | Yes |
| *Trichoderma viride* (HF3MP) | 6.03 | 8.2 | Yes |
| *Trichoderma viride* (HF3TV) | 6.02 | 8.5 | Yes |
| Agar Plug Control | 0.00 | 8.8 | N/A |

### 3.2. Co-Culture Interaction Analysis

Fungal strains demonstrating robust survivability in regolith-supplemented media were subsequently paired with the diazotrophic cyanobacterium *Anabaena* sp. (UTEX 2576) to evaluate their capacity to form stable autotrophic–heterotrophic consortia under regolith-relevant conditions. *Anabaena* spp. have been widely investigated as candidate organisms for ISRU due to their ability to grow using Martian regolith as a resource and to sustain metabolism under nutrient-limited conditions (Verseux et al. 2016; Macário et al. 2022; Arribas Tiemblo et al. 2025).

Cultures were grown in modified Bold's Basal Medium (BBM) amended with a pH buffer (MOPS) and sterilized MMS-2, relying exclusively on light, air, and inorganic nutrients. To systematically assess metabolic interdependence, four nutrient regimes were tested: C+N+, C+N−, C−N+, and C−N−, where cellulose and $(NH_4)_2SO_4$ served as representative supplemental carbon and nitrogen sources, respectively. Ammonium is a preferred nitrogen source for many cyanobacteria (Canizales et al., 2021) and can be metabolically assimilated with low energetic cost, while low concentrations of sulfate, potentially obtainable from Martian gypsum deposits, have been reported to support cyanobacterial growth (Olsson-Francis and Cockell, 2010). Cellulose was chosen as a model organic carbon source because it is readily utilized by many filamentous fungi (Antonov et al., 2016), including several strains tested here. These supplements were included solely as experimental controls; the C−N− condition was the focus of this study, which was designed to approximate closed-loop, resource-limited growth relevant to Martian environments.

Macroscopic observations revealed distinct visual signatures of microbial activity. The base medium exhibited a reddish coloration due to MMS-2; however, dense cyanobacterial growth produced a visible green to dark-green shift in culture wells (Figure S2), enabling direct qualitative assessment of photoautotrophic proliferation. Abundant gas bubbles were observed in cyanobacterial monocultures and co-cultures, consistent with active oxygen evolution during photosynthesis. In parallel, fungal monocultures and several co-culture systems developed visible white mycelial aggregates, indicating sustained heterotrophic growth. In certain pairings, such as those involving *P. brefeldianum*, pronounced mycelial clustering was evident.

The pH of the abiotic control (modified BBM supplemented with MMS-2 and MOPS, without microbial inoculation) was 8.37 at the start of the experiment. Over the 28-day incubation period, microbial treatments exhibited modest but measurable pH shifts in both directions (Table 2). Final pH values ranged from 8.29 to 8.50. The highest pH (8.50) was observed in two co-culture systems containing *P. stipitatus* and *A. niger*, respectively, whereas the lowest pH (8.29) occurred in the co-culture containing *T. viride* (HF3TV). Although the magnitude of change was limited, these shifts indicate active microbial modulation of carbonate–bicarbonate equilibria under regolith-relevant alkaline conditions, likely reflecting the balance between photosynthetic $CO_2$ uptake and fungal metabolic activity.

Fluorescence microscopy images (Figure 4) revealed distinct differences in cyanobacterial viability among the C−N− co-culture systems after 28 days of incubation. In co-cultures containing *A. restrictus* (F1), *P. atramentosum* (F2), *A. malodoratus* (F4), and *P. brefeldianum* (F5), only sparse live cyanobacterial cells were detected, indicating limited compatibility under nutrient-deprived conditions. In contrast, the remaining C−N− co-cultures displayed abundant viable cyanobacterial cells closely associated with and interwoven among fungal hyphae. This spatial co-localization suggests active physiological coupling between partners and is consistent with the establishment of metabolically integrated interactions under carbon- and nitrogen-limited conditions.

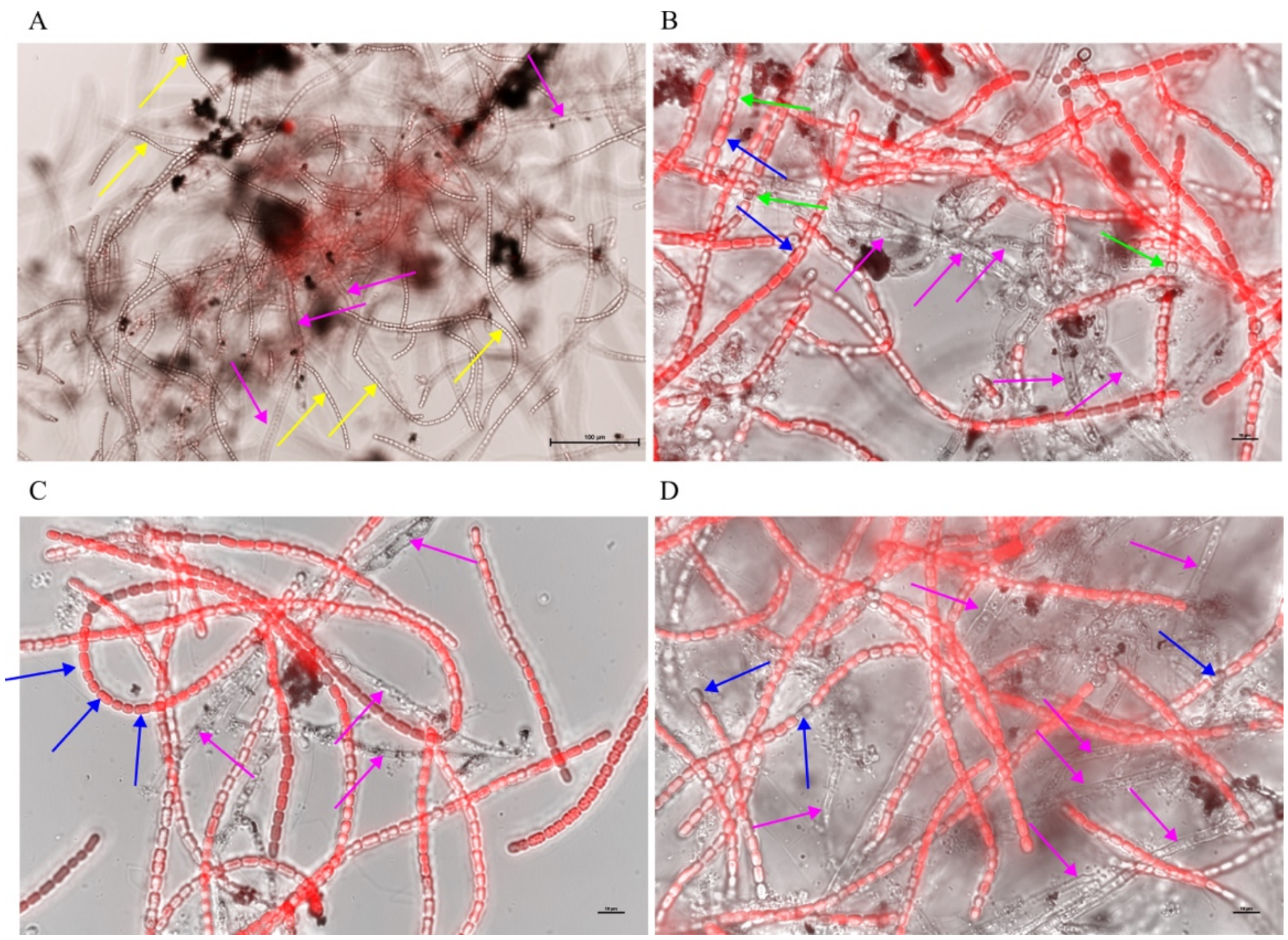

**FIGURE 4.** Microscopic images of the C-N- co-culture wells whose fungal component is (A) F5: *P. brefeldianum,* (B) F9: *T. viride* (HF3MP), (C) F8: *T. viride* (HF3TV), and (D) F7: *T. reesei*. Red-colored fluorescing cells are live cyanobacterial cells, yellow arrows indicate dead cyanobacterial cells, purple arrows indicate fungal cells, green arrows indicate heterocysts, and blue arrows indicate akinetes. There were almost no live cyanobacterial cells in (A), whereas cyanobacteria exhibited robust growth in (B) to (D). The scale bar is 100 μm in (A) and 10 μm in (B) to (D).

**Table 2.** The pH of each well after 28-day incubation.

| Label | Fungal Strain | Monoculture | | | | Co-Culture | | | |
|---|---|---|---|---|---|---|---|---|---|
| | | C+N+ | C+N- | C-N+ | C-N- | C+N+ | C+N- | C-N+ | C-N- |
| F1 | *A. restrictus* | 8.45 | 8.40 | 8.42 | 8.42 | 8.42 | 8.34 | 8.33 | 8.46 |
| F2 | *P. atramentosum* | 8.41 | 8.36 | 8.34 | 8.38 | 8.41 | 8.35 | 8.33 | 8.43 |
| F3 | *P. stipitatus* | 8.41 | 8.40 | 8.34 | 8.40 | 8.40 | 8.34 | 8.50 | 8.47 |
| F4 | *A. malodoratus* | 8.41 | 8.42 | 8.41 | 8.40 | 8.48 | 8.37 | 8.32 | 8.42 |
| F5 | *P. brefeldianum* | 8.37 | 8.34 | 8.31 | 8.38 | 8.44 | 8.36 | 8.32 | 8.43 |
| F6 | *A. niger* | 8.36 | 8.33 | 8.31 | 8.36 | 8.39 | 8.44 | 8.47 | 8.50 |
| F7 | *T. reesei* | 8.49 | 8.41 | 8.40 | 8.41 | 8.39 | 8.33 | 8.33 | 8.39 |
| F8 | *T. viride* (HF3TV) | 8.37 | 8.36 | 8.36 | 8.35 | 8.39 | 8.32 | 8.29 | 8.36 |
| F9 | *T. viride* (HF3MP) | 8.35 | 8.34 | 8.31 | 8.33 | 8.36 | 8.39 | 8.41 | 8.43 |
| F10 | *A. flavus* | 8.36 | 8.36 | 8.36 | 8.37 | 8.40 | 8.31 | 8.31 | 8.39 |

### 3.3. Quantitative Assessment of Microbial Growth and Metabolic Activity

Phycocyanin fluorescence measurements after 28 days of incubation are presented in Figure 5A. As expected, cyanobacterial growth was maximal under nutrient-replete conditions (C+N+). It should be noted that the co-cultures were inoculated with only half the initial cyanobacterial cell density compared to the cyanobacterial monoculture. Despite this lower starting concentration, cyanobacteria in the C+N+ co-culture wells generally achieved growth levels comparable to, or exceeding, those observed in axenic culture, indicating that the presence of compatible fungal partners did not inhibit cyanobacterial proliferation under nutrient-sufficient conditions. Under C−N− conditions, phycocyanin autofluorescence was markedly reduced in co-cultures containing *A. restrictus* (F1), *P. atramentosum* (F2), *A. malodoratus* (F4), and *P. brefeldianum* (F5), consistent with the limited cyanobacterial viability observed microscopically. In contrast, the remaining C−N− co-cultures exhibited cyanobacterial growth levels comparable to or exceeding those in cyanobacterial monoculture under the same nutrient limitation.

These results demonstrate that fungal partner identity is a critical determinant of co-culture performance under nutrient-poor conditions. Only a subset of strains supports sustained cyanobacterial growth in the absence of external carbon and nitrogen inputs, indicating that mutualistic stabilization depends on specific interspecies compatibility rather than generic co-cultivation. This differential performance underscores the importance of strain selection in engineering metabolically self-sustaining consortia for regolith-based biomineral production.

Resazurin assay results after 28 days of incubation are presented in Figure 5B. Overall, fungal metabolic activity in co-culture wells, particularly under C−N− conditions, was substantially higher than in corresponding fungal monocultures. This trend suggests that nutrient-poor conditions may promote metabolic interdependence and favor the establishment of cooperative interactions. However, the balance between partners varied markedly depending on fungal identity. In the C−N− co-culture wells containing F1 (*A. restrictus*), F2 (*P. atramentosum*), and F4 (*A. malodoratus*), resazurin signals were high while phycocyanin fluorescence was low. This inverse pattern indicates vigorous fungal metabolism accompanied by severely reduced cyanobacterial viability, consistent with competitive dominance rather than stable mutualism. In contrast, C−N− co-cultures containing F3, F6, F7, F8, F9, and F10 exhibited elevated signals in both resazurin and phycocyanin assays relative to their respective monocultures. In these systems, both fungal and cyanobacterial partners benefited from co-cultivation, indicating successful metabolic coupling under nutrient limitation.

As summarized in Table 3, two general interaction patterns emerged under C−N− conditions. First, excessive fungal proliferation often corresponded to suppressed cyanobacterial growth (F1, F2, F4), suggesting competitive resource sequestration or physical overgrowth. Second, moderate fungal growth more frequently coincided with enhanced cyanobacterial performance (F7–F10), reflecting balanced metabolic exchange. Notably, exceptions to both patterns were observed (e.g., F3 and F6), underscoring that mutualistic stability depends not simply on growth magnitude but on the qualitative nature of metabolic integration between partners. These results demonstrate that successful regolith-based co-cultures require deliberate strain pairing to achieve metabolic complementarity rather than competitive dominance.

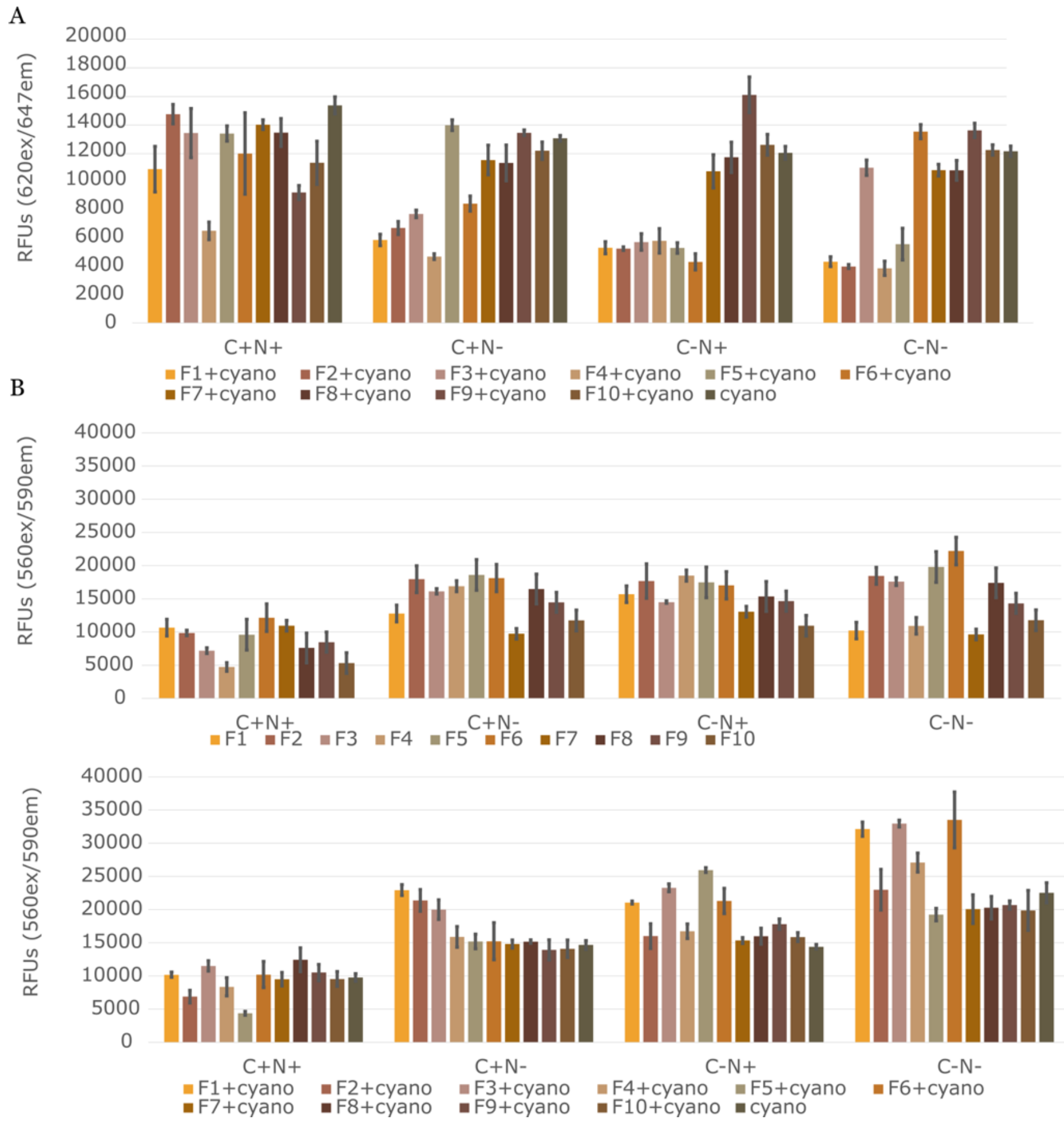

**FIGURE 5.** Results of (a) phycocyanin assay and (b) resazurin assay after 28-day incubation.

**Table 3.** The cyanobacterial and fungal growth in C-N- co-culture wells.

| | Excellent fungal growth | Moderate fungal growth |
|---|---|---|
| Poor cyanobacterial growth | F1: *A. restrictus* | F5: *P. brefeldianum* |
| | F2: *P. atramentosum* | |
| | F4: *A. malodoratus* | |
| Good cyanobacterial growth | F3: *T. stipitatus* | F7: *T. reesei* |
| | F6: *A. niger* | F8: *T. viride* (HF3TV) |
| | | F9: *T. viride* (HF3MP) |
| | | F10: *A. flavus* |

Fungal viability after 28 days of incubation was further evaluated by plating samples onto selective medium (Figure 6A). While this approach provides only a semi-quantitative estimate of viable fungal cells, it confirms the sustained presence of metabolically active fungi across conditions. Colony surface coverage exceeded 25% in all C−N− co-culture wells and 15% in all C+N+ co-culture wells, indicating substantial fungal survival even under carbon- and nitrogen-limited conditions. Although plating does not capture total biomass or account for non-culturable cells, the consistently high coverage supports the conclusion that fungal partners remained robust throughout the experimental period. More quantitative viability assays will be implemented in future studies to refine these estimates.

Importantly, microscopic observations revealed that fine granular particles of the Martian regolith simulant were physically bound together by fungal hyphae and associated biomineral deposits (Figure 6B). The representative image was obtained from the C−N− co-culture containing T. *viride* (HF3MP). Hyphal networks bridged adjacent regolith particles, forming filamentous, mineralized scaffolds that entrapped and stabilized the granular substrate. This early-stage consolidation provides direct evidence that the engineered microbial consortia not only persist in regolith-containing media but also actively reorganize and reinforce the surrounding material environment. Such biologically mediated particle binding represents a critical initial step toward consortia-driven mineral consolidation in granular systems and supports the feasibility of living, self-strengthening materials under nutrient-limited conditions.

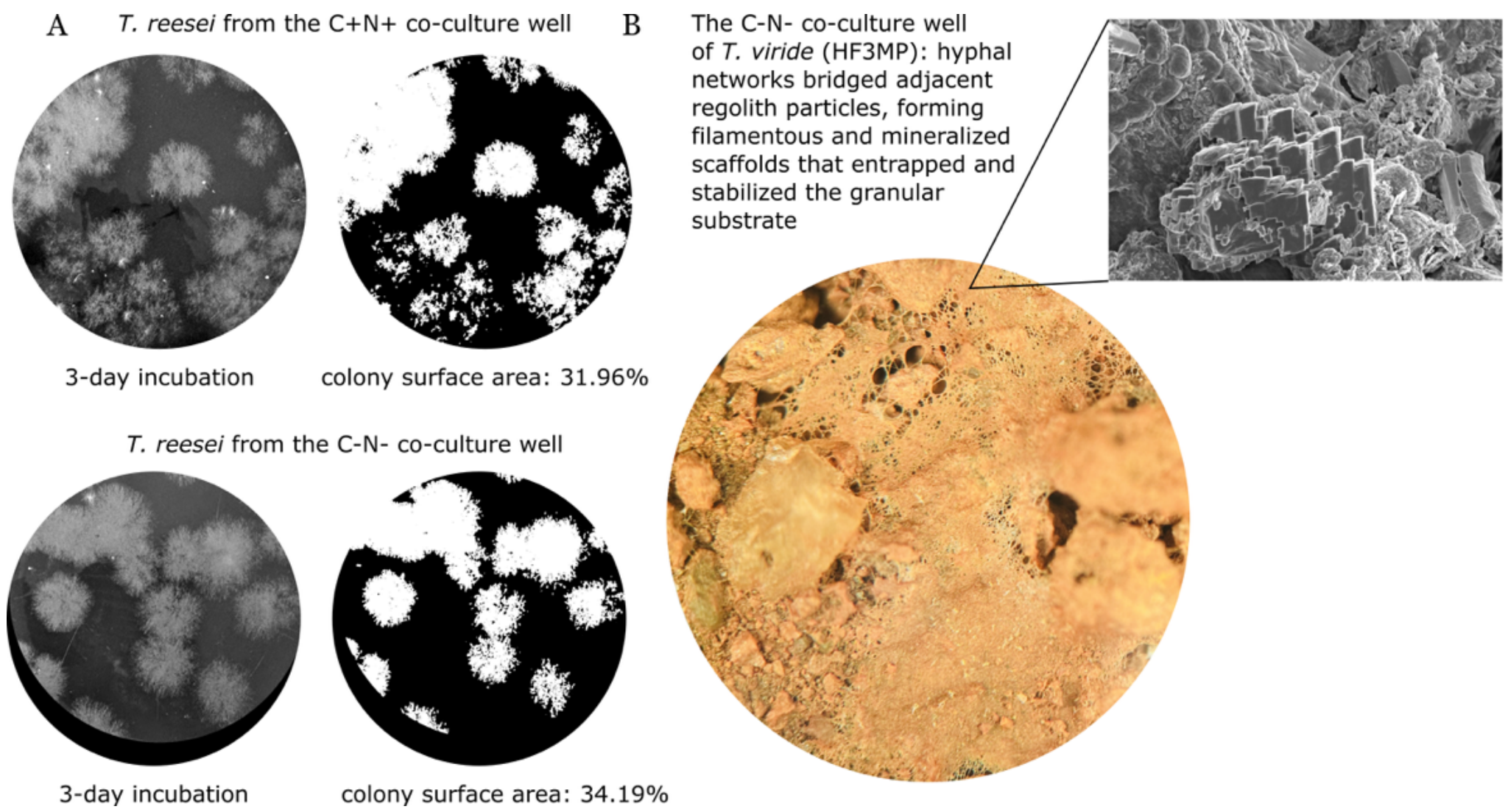

**FIGURE 6.** (A) Fungal plating on selective medium after 28-day incubation. (B) Martian regolith simulants were bonded together by the microbial filaments and the biominerals they produced.

### 3.4. Metabolomics Analysis

To investigate how *Anabaena* sp. metabolically supports its fungal partner under carbon- and nitrogen-limited conditions, two high-performing co-culture systems, i.e., *Anabaena* sp. with *T. reesei* and *T. viride* (HF3MP), were selected for untargeted metabolomic analysis. Untargeted metabolomics enables comprehensive profiling of small molecules without prior selection of target metabolites, providing a systems-level view of metabolic reprogramming within symbiotic communities. Metabolite extraction and analysis were performed using gas chromatography–mass spectrometry (GC–MS), a sensitive and widely established platform for detecting volatile and semi-volatile compounds (Papadimitropoulos et al. 2018).

To evaluate global metabolic shifts and identify metabolites contributing to group differentiation, partial least squares–discriminant analysis (PLS-DA) was performed. This supervised multivariate method projects high-dimensional metabolomic data into latent variables that maximize separation between experimental groups (Szymańska et al. 2012). The resulting score plots visualize clustering patterns among monocultures and co-cultures, enabling assessment of metabolic divergence and reproducibility. Tight clustering within replicates and clear separation between groups indicate robust and reproducible metabolic distinctions associated with co-culture interactions. Using this analytical framework, we identified coordinated metabolic changes underlying autotrophic–heterotrophic interactions under nutrient-limited conditions, providing insight into the redistribution of carbon and nitrogen fluxes within the engineered consortia.

As shown in Figure 7, the PLS-DA score plot reveals clear separation among the five sample groups, indicating pronounced differences in their metabolite compositions. Each datapoint represents an individual biological replicate, and the tight clustering within groups demonstrates high reproducibility and experimental consistency. Distinct, non-overlapping clusters were observed for the *Anabaena* monoculture (Ana), the fungal monocultures *T. reesei* (TR) and *T. viride* (TV), and the two co-culture systems (Ana.TR and Ana.TV). Notably, the two co-culture groups clustered in close proximity to one another while remaining clearly separated from all monocultures. This pattern suggests that co-cultivation gives rise to a shared metabolomic state that is distinct from either partner grown alone, consistent with coordinated metabolic reprogramming. The convergence of metabolite profiles indicates that interspecies interactions drive the emergence of an integrated metabolic network, which likely contributes to the enhanced stability and cooperative functionality observed in the engineered consortia.

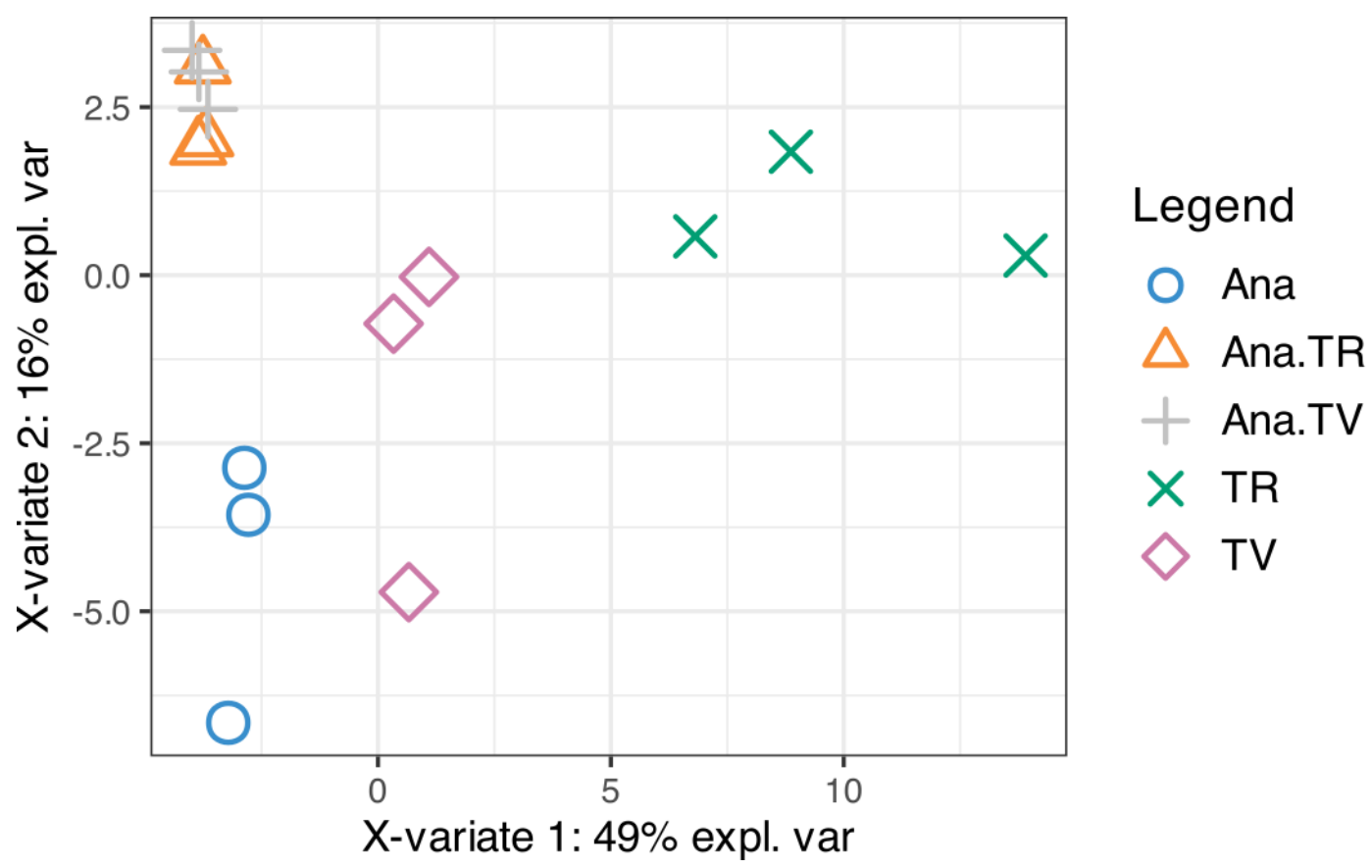

**FIGURE 7.** PLS-DA score plot showing the separation of sample groups based on their metabolite profiles. Each datapoint represents a sample, and clustering indicates similarity within groups.

Across all treatments, a total of 64 metabolites were detected. Variable importance in projection (VIP) scores were calculated for each metabolite based on the PLS-DA model (Figure S3A). VIP scores quantify the contribution of individual metabolites to group separation, with values greater than 1.0 commonly considered significant in metabolomic analyses (Nishanth and Prasanna, 2022). Of the 64 metabolites identified, 30 exhibited higher relative abundance in the *Anabaena* monoculture (Ana) compared to the fungal monocultures, i.e., *T. reesei* (TR) and *T. viride* (TV), indicating that the cyanobacterial metabolic profile was distinct and comparatively enriched in specific compounds. In contrast, the two co-culture systems (Ana.TR and Ana.TV) displayed broadly similar metabolite expression patterns, further supporting the convergence of metabolic states observed in the PLS-DA clustering.

To further characterize treatment-specific metabolic shifts, differential expression analysis was performed. Metabolites were defined as differentially expressed if they satisfied two criteria (Antunes et al. 2019): an absolute fold change ≥ 1 ($|\log_2 \text{ratio}| \geq 1$) and a p-value $< 0.05$. Based on these thresholds, 43 of the 64 detected metabolites were identified as differentially expressed in at least one pairwise comparison (Figure S3B). These results indicate extensive metabolic remodeling across monoculture and co-culture conditions, highlighting the magnitude of biochemical reorganization associated with interspecies interactions under nutrient-limited, regolith-relevant conditions.

Comparison of differential expression patterns revealed markedly greater downregulation in the Ana.TR/TR and Ana.TV/TV contrasts than in the Ana.TR/Ana and Ana.TV/Ana contrasts. Relative to the *Anabaena* monoculture (Ana), the Ana.TR co-culture exhibited two upregulated and seven downregulated metabolites. In contrast, when compared to the *T. reesei* monoculture (TR), the Ana.TR co-culture showed 27 downregulated metabolites. A similar trend was observed in the Ana.TV system: relative to Ana, the co-culture displayed one upregulated and four downregulated metabolites, whereas comparison with the *T. viride* monoculture (TV) revealed 33 downregulated metabolites. This asymmetry in metabolic shifts indicates that the co-culture metabolic state more closely resembles that of *Anabaena* than that of the fungal monocultures. In other words, *Anabaena* appears to retain functional dominance within the engineered consortia, while the fungal partners undergo substantial metabolic downregulation. Such widespread suppression of fungal-associated metabolites suggests a shift toward metabolic dependency on cyanobacterially derived carbon and nitrogen compounds. These findings are consistent with a functional organization in which the photoautotrophic partner drives primary metabolism and nutrient provision, while the heterotrophic partner adapts to a resource-sharing regime under nutrient-limited conditions.

Sugars and their derivatives, amino acids and nitrogenous bases, and organic acids were among the most significantly affected metabolite classes, all of which are central to carbon and nitrogen metabolism. In the sugar category, comparison with the *Anabaena* monoculture (Ana) revealed that mannitol and cellobiose were significantly upregulated in the Ana.TR co-culture, whereas glucose and glycerol were downregulated. In the Ana.TV co-culture, glucose was likewise reduced relative to Ana. Mannitol and cellobiose are known to play important roles in osmotic regulation, carbon storage, and oxidative stress mitigation (Meena et al. 2015; Ruijter et al. 2003), suggesting enhanced protective responses and carbon allocation within the engineered consortia. Conversely, the reduction in glucose and glycerol likely reflects active consumption or metabolic redirection into downstream pathways that support interspecies nutrient exchange and overall co-culture stability (Fritts et al. 2021). These coordinated shifts in carbohydrate metabolism further support the establishment of functional metabolic integration between the autotrophic and heterotrophic partners.

In the case of nitrogenous metabolites, comparison with the *Anabaena* monoculture (Ana) revealed that aspartic acid, nicotinic acid, and 5-aminovaleric acid were consistently upregulated in both Ana.TR and Ana.TV co-cultures. Aspartic acid serves as a key node in nitrogen assimilation and amino acid biosynthesis, while nicotinic acid is central to $NAD^+$ metabolism and cellular redox balance. The coordinated upregulation of these metabolites suggests enhanced nitrogen flux and metabolic coordination within the engineered consortia. These findings support the hypothesis that *Anabaena* sp., through its diazotrophic capacity, contributes significantly to nitrogen provisioning in the co-culture, potentially supplying amino acids and other nitrogen-containing metabolites to its fungal partner (Zhang et al. 2018; Chen et al. 2025). Such metabolic reallocation is consistent with a transition toward interdependent nutrient exchange and stable autotrophic–heterotrophic integration.

In the case of organic acids, multiple intermediates of the tricarboxylic acid (TCA) cycle, including fumaric acid, malic acid, α-ketoglutaric acid, oxaloacetic acid, succinic acid, and citric acid, were upregulated in both Ana.TR and Ana.TV co-cultures relative to the Ana monoculture. The coordinated elevation of TCA cycle intermediates suggests enhanced central carbon metabolism in *Anabaena* under co-culture conditions, potentially increasing the pool of organic acids available as transferable carbon substrates for the fungal partner (Flores and Herrero 2005; Jin et al. 2022). In addition, aromatic acids such as 4-methylbenzoic acid and shikimic acid were elevated, indicating a shift toward aromatic compound metabolism. This metabolic reorientation may reflect stress adaptation, redox balancing, or interkingdom signaling processes that contribute to stabilization of the co-culture system (Matsuyama et al., 2009).

While sustained activity under C−N− conditions is consistent with autotrophic carbon fixation and diazotrophic nitrogen fixation by the cyanobacterial partner, direct quantification of these processes was beyond the scope of the present study. Future work incorporating isotopic labeling approaches (e.g., $^{13}CO_2$ and $^{15}N_2$) will be necessary to directly quantify carbon and nitrogen fixation rates and to further resolve interspecies nutrient exchange mechanisms.

**4. Conclusions, Limitations, and Outlook**

In this study, we constructed and evaluated six engineered autotrophic–heterotrophic consortia composed of the diazotrophic cyanobacterium *Anabaena* sp. and filamentous fungi (*P. stipitatus, A. niger, T. reesei, T. viride* [HF3TV], *T. viride* [HF3MP], and *A. flavus*). These consortia demonstrated sustained growth in the presence of Martian regolith simulant under light-driven, inorganic nutrient conditions, without supplementation of organic carbon or nitrogen. Metabolomic analyses revealed coordinated shifts in carbon- and nitrogen-associated metabolites in co-culture relative to monoculture, consistent with metabolic integration and nutrient exchange. These findings establish a biological foundation for consortia-driven, autonomous biomineralization and material formation under regolith-relevant conditions.

It is important to emphasize that the present study was conducted under Earth-like laboratory conditions using Martian regolith simulants and does not replicate true Martian environmental conditions. The transparent incubator shown in Figure 1 represents a conceptual system design intended to provide a controlled, enclosed environment for future deployment; however, its performance under Martian conditions has not been experimentally validated. If the experiments are performed under actual Martian stressors (Jakosky et al., 2018), including low temperature, desiccation, reduced pressure, and ionizing radiation, microbial metabolic rates are expected to be substantially reduced or potentially halted, which would directly impact biomineralization kinetics and overall system performance. In addition, the chemical composition of Martian regolith, including the presence of highly oxidizing species such as perchlorates (Hecht et al., 2009), may further influence microbial activity and metabolic pathways. These limitations highlight the need for future studies that integrate realistic environmental stressors and geochemical complexity to rigorously evaluate system viability under in situ conditions.

Although Mars provides an extreme testbed, the underlying principles demonstrated here are not limited to extraterrestrial environments. Systems capable of coupling metabolic activity with material formation under minimal-input conditions are broadly relevant to manufacturing in extreme or resource-constrained environments, including remote terrestrial locations, arid regions, and infrastructure-limited settings. In this context, engineered microbial consortia offer a platform for distributed, low-energy material production in which structure formation and material consolidation emerge from biologically driven processes. This work highlights a broader paradigm: that future manufacturing may increasingly rely on living systems capable of adapting to and transforming challenging environments, extending beyond space exploration to sustainable and resilient materials production on Earth.

**5. Materials and Methods**

**5.1. Survivability Screening**

To evaluate fungal tolerance to regolith-relevant conditions, survivability assays were conducted using Enhanced Mojave Mars Simulant (MMS-2; The Martian Garden, Austin, TX, USA). Growth medium was prepared by incorporating sterilized MMS-2 into PDA at a concentration of 0.1 g $mL^{-1}$ prior to solidification. Control media consisted of PDA without regolith simulant.

Fungal inoculation was performed using a 6-mm-diameter mycelial plug excised from the actively growing margin of a 7-day-old culture. Plugs were centrally placed onto 30-mm Petri dishes containing 15 mL of prepared medium. Plates were incubated at 25 °C under ambient light conditions for 14 days. Each condition was performed in triplicate. Sterile PDA plugs without fungal inoculum served as negative controls to confirm the absence of contamination.

Following incubation, radial growth was measured to determine colony expansion rates. To assess potential regolith-induced pH modulation, agar media from each plate were melted and homogenized, and pH measurements were obtained using an Orion double-junction pH electrode. Three independent measurements were recorded per sample and averaged for analysis.

**5.2. Co-Culture Interaction Analysis**

Fungal spores were harvested in 20% glycerol and stored at −80 °C until use. Prior to inoculation, frozen spores were thawed at 25 ± 2 °C for 1 h. The diazotrophic cyanobacterium (*Anabaena* sp.) was propagated in BG11 medium for two weeks under standard growth conditions before experimental setup. A modified BBM served as the base medium for both monoculture and co-culture experiments. Cyanobacterial monocultures were inoculated to an initial optical density ($OD_{720}$) of 0.25. Fungal monocultures were inoculated at a final concentration of $5 \times 10^5$ spores $mL^{-1}$. For co-culture conditions, equal volumes of cyanobacterial suspension and fungal spore suspension were mixed prior to inoculation, yielding a total inoculum volume of 150 µL per well in sterile 24-well plates. All cultures were grown under standing conditions with illumination at 100 µmol photons $m^{-2}$ $s^{-1}$ using a 14:10 h light:dark cycle. Temperature was maintained at 25 ± 1 °C with 60% relative humidity. Cultures were exposed to filter-sterilized air throughout the incubation period.

To evaluate the influence of exogenous carbon and nitrogen availability, the base medium was supplemented with cellulose as carbon source and ammonium sulfate $(NH_4)_2SO_4$ as nitrogen source, respectively. Four nutrient conditions were systematically tested for each monoculture and co-culture: C+N+, C+N−, C−N+, and C−N−. The C+ condition corresponded to cellulose supplementation at 0.12 g per 100 mL, while the N+ condition corresponded to $(NH_4)_2SO_4$ supplementation at 2.94 mM. Each well contained 2 mL of growth medium. Sterilized MMS-2 was added to a final concentration of 0.1 g $mL^{-1}$ to simulate regolith-relevant conditions. The inert buffering agent MOPS (20 mM; Fisher Scientific, Pittsburgh, PA, USA) was included to stabilize pH throughout incubation. Control wells containing MMS-2 and MOPS but no microbial inoculum were prepared to account for abiotic background effects. All experimental conditions were conducted in triplicate.

**5.3. Quantitative Assessment of Microbial Growth and Metabolic Activity**

Microbial growth and metabolic activity were monitored longitudinally throughout the 28-day incubation period. Fluorescence microscopy was performed using excitation wavelengths between 460–550 nm to visualize cyanobacterial autofluorescence (primarily from phycobiliproteins) and to qualitatively assess spatial interactions between cyanobacteria and fungal hyphae. The pH of each well was measured periodically to evaluate shifts in carbonate chemistry and metabolic acid–base dynamics. To quantitatively assess viability and metabolic activity, three complementary methods were employed: resazurin reduction assay, phycocyanin fluorescence assay, and fungal plating on selective medium.

The resazurin reduction assay was used to estimate viable biomass in each well after 28 days of incubation. The assay is based on the reduction of oxidized, blue, non-fluorescent resazurin to resorufin, a red, highly fluorescent product, by redox-active metabolic processes (e.g., respiratory and other oxidoreductase activity) in viable cells. The amount of resorufin produced is directly proportional to the number of metabolically active cells and can be quantified fluorometrically. At the end of the incubation period, 100 µL of homogenized culture from each well was withdrawn in duplicate and transferred to a 96-well microplate. Resazurin working solution was pre-warmed to 37 °C to ensure complete dissolution, and 20 µL was added to each well. Plates were incubated for 4 h at 25 ± 2 °C under dark conditions to prevent photodegradation. Fluorescence was measured using a BioTek Synergy H1 hybrid spectrophotometer (Agilent Technologies, Santa Clara, CA, USA) with excitation at 560 nm and emission at 590 nm. Relative fluorescence units (RFUs) were recorded.

Cyanobacterial growth was quantified by measuring the autofluorescence of phycocyanin, a photosynthetic pigment–protein complex unique to cyanobacteria and absent in fungi. Because phycocyanin degrades rapidly in dead cells through decomposition or photodegradation, fluorescence intensity serves as a proxy for live cyanobacterial biomass. After 28 days of incubation, fluorescence area scans were performed directly on the 24-well plates using the BioTek Synergy H1 spectrophotometer. Measurements were acquired with excitation at 620 nm and emission at 647 nm. The average RFU for each well was calculated following background subtraction from regolith-containing control wells without cyanobacteria.

Viable fungal populations were further assessed through plating on selective medium. After 28 days of incubation, 100 µL of homogenized culture from each well was spread evenly onto modified PDA plates containing chloramphenicol (0.25 g/L) to suppress cyanobacterial growth. Plates were incubated at 25 ± 2 °C for 3 days. Colony expansion was quantified by estimating surface area coverage. Plates were photographed under standardized lighting conditions, and images were converted to binary format using image-processing software. The ratio of white pixels (fungal colony area) to total pixels was calculated as a measure of relative fungal biomass.

**5.4. Metabolomics Analysis**

To investigate the metabolic mechanisms underlying nutrient exchange in autotrophic–heterotrophic consortia under carbon- and nitrogen-limited conditions, two high-performing co-culture systems, i.e., *Anabaena* sp. paired with *T. reesei* and *T. viride* (HF3MP), were selected for untargeted metabolomic profiling. Supernatants from 28-day-old triplicate cultures grown under C−N− conditions were rapidly flash-frozen in liquid nitrogen and stored at −80 °C until extraction to preserve metabolite integrity.

Metabolite extraction was performed following a modified methanol-based protocol (Hayer et al. 2024). Briefly, 1 mL of culture supernatant was mixed with 100% ice-cold methanol (−20 °C) to precipitate proteins and quench metabolic activity. Pinitol (16 µL of 0.04 µg/µL) was added as an internal standard to monitor extraction consistency. Samples were vortexed, incubated on ice, and centrifuged to remove precipitated material. A 20 µL aliquot of clarified supernatant was transferred to fresh tubes and dried under vacuum.

For derivatization, dried extracts were treated with N-methyl-N-(trimethylsilyl)trifluoroacetamide (MSTFA) containing 1% trimethylchlorosilane (TMCS) (Thermo Fisher Scientific, Pittsburgh, PA, USA) and incubated at 37 °C for 30 min with agitation (1000 rpm) to convert polar metabolites into volatile trimethylsilyl derivatives suitable for gas chromatography. Following centrifugation (16,000 × g, 10 min, room temperature), supernatants were transferred to glass autosampler vials for GC–MS analysis.

GC–MS analysis was performed using an Agilent 7890B gas chromatograph coupled to a 5977A mass selective detector (Agilent Technologies, Santa Clara, CA, USA) equipped with an HP-5MS column and helium as the carrier gas. Samples were injected in split mode (1:5) at 260 °C. The oven temperature was programmed from 60 °C to 325 °C at a rate of 10 °C $min^{-1}$. A C10–C25 alkane retention index standard was used for chromatographic calibration. Quality control (QC) pooled samples were analyzed four times throughout the analytical sequence to monitor instrument stability and enable cross-batch normalization. Due to interference from the MOPS buffer used to stabilize pH in regolith-containing media, the internal standard was not detectable after sample concentration and was therefore excluded from normalization procedures.

Raw data were processed using MS-DIAL (v4.48) for peak deconvolution, alignment, and metabolite identification against local and public retention index (RI) spectral libraries, including the Fiehn, RIKEN, and MoNA databases. Only metabolites with RI similarity >95% were retained for further analysis. Peak areas were normalized using locally weighted scatterplot smoothing (LOWESS) to correct for systematic batch effects. LOWESS is a non-parametric regression approach that reduces technical variation without imposing assumptions about data distribution.

Multivariate statistical analysis was conducted in R. PLS-DA was performed using the PLS package to evaluate separation among treatments. Model robustness was assessed using cross-validation and permutation testing to evaluate model stability and avoid overfitting. VIP scores were calculated to identify metabolites contributing most strongly to group discrimination. Differential expression analysis was performed using Student's t-tests to compare co-culture and monoculture conditions. Metabolites were considered differentially expressed if they satisfied two criteria: absolute $\log_2$ fold change ≥ 1 and $p < 0.05$. Heatmaps of normalized metabolite abundance and $\log_2$ fold changes were generated using the heatmap package in R.

**Author contributions**

N.R. led experimental work, data analysis, and drafted the initial manuscript. E.C.C. contributed to experimental work, data analysis, and manuscript drafting and editing. K.S. contributed to experimental work and methodology development. R.A.W. contributed to resources, supervision, and manuscript editing. Y.H. contributed to project conceptualization and manuscript editing. C.J. led project conceptualization, supervision, funding acquisition, data analysis, and manuscript drafting and editing.

**Acknowledgments**

This study was funded by the NASA Innovative Advanced Concepts (NIAC) program of the National Aeronautics and Space Administration (NASA) under the grant number of 80NSSC23K0584. Erin C. Carr was supported by Postdoctoral Research Fellowships in Biology (PRFB) program of National Science Foundation (NSF) under the award number of 2209217. Proteomics & Metabolomics Facility (RRID:SCR_021314), Nebraska Center for Biotechnology at the University of Nebraska-Lincoln is thanked for the metabolomics analysis. Dr. Steven Harris at the Department of Plant Pathology, Entomology and Microbiology at Iowa State University is thanked for useful discussion and providing some of the fungal strains.

**Conflicts of Interest**

The authors declare no conflicts of interest.

**Data Availability Statement**

The data that support the findings of this study are available from the corresponding author upon reasonable request.

**Supplementary Information**

# Weaving Life into Regolith: Engineered Autotrophic–Heterotrophic Consortia for Autonomous Biofabrication from Granular Feedstocks

Nisha Rokaya [a], Erin C. Carr [b], Kumar Shrestha [c], Richard A. Wilson [a], Yong Huang [d], Congrui Jin [e*]

[a] Department of Plant Pathology, University of Nebraska–Lincoln, Lincoln, NE 68583

[b] School of Biological Sciences, University of Nebraska–Lincoln, Lincoln, NE 68588

[c] Agricen Science, Pilot Point, TX 76258

[d] Department of Mechanical and Aerospace Engineering, University of Florida, Gainesville, FL 32611

[e] Department of Engineering Technology and Industrial Distribution, Texas A&M University, College Station, TX 77843

[*] Corresponding author: Congrui Jin; E-mail addresses: jincongrui@tamu.edu

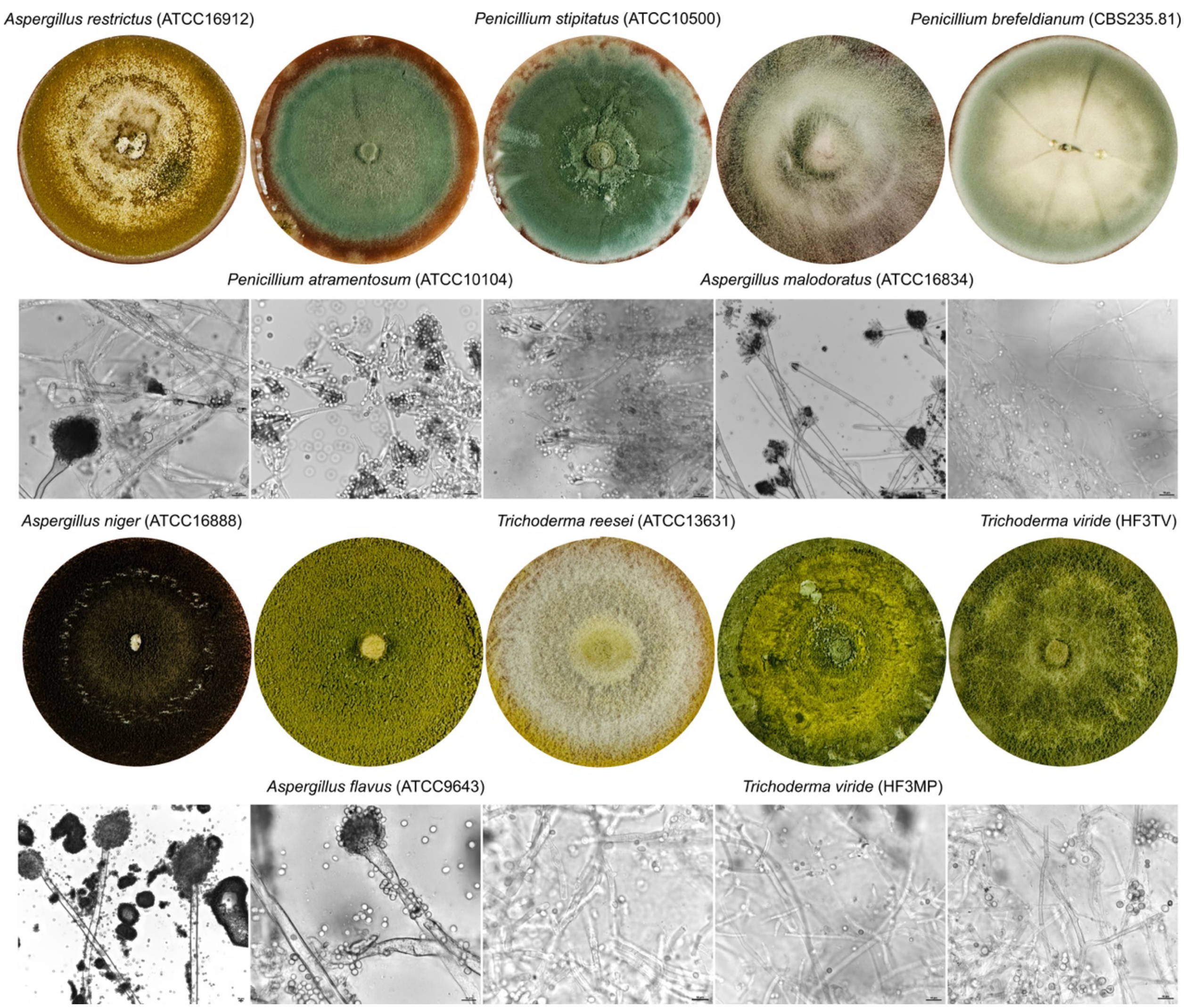

**FIGURE S1**. Fungal strains selected based on the results of survivability testing.

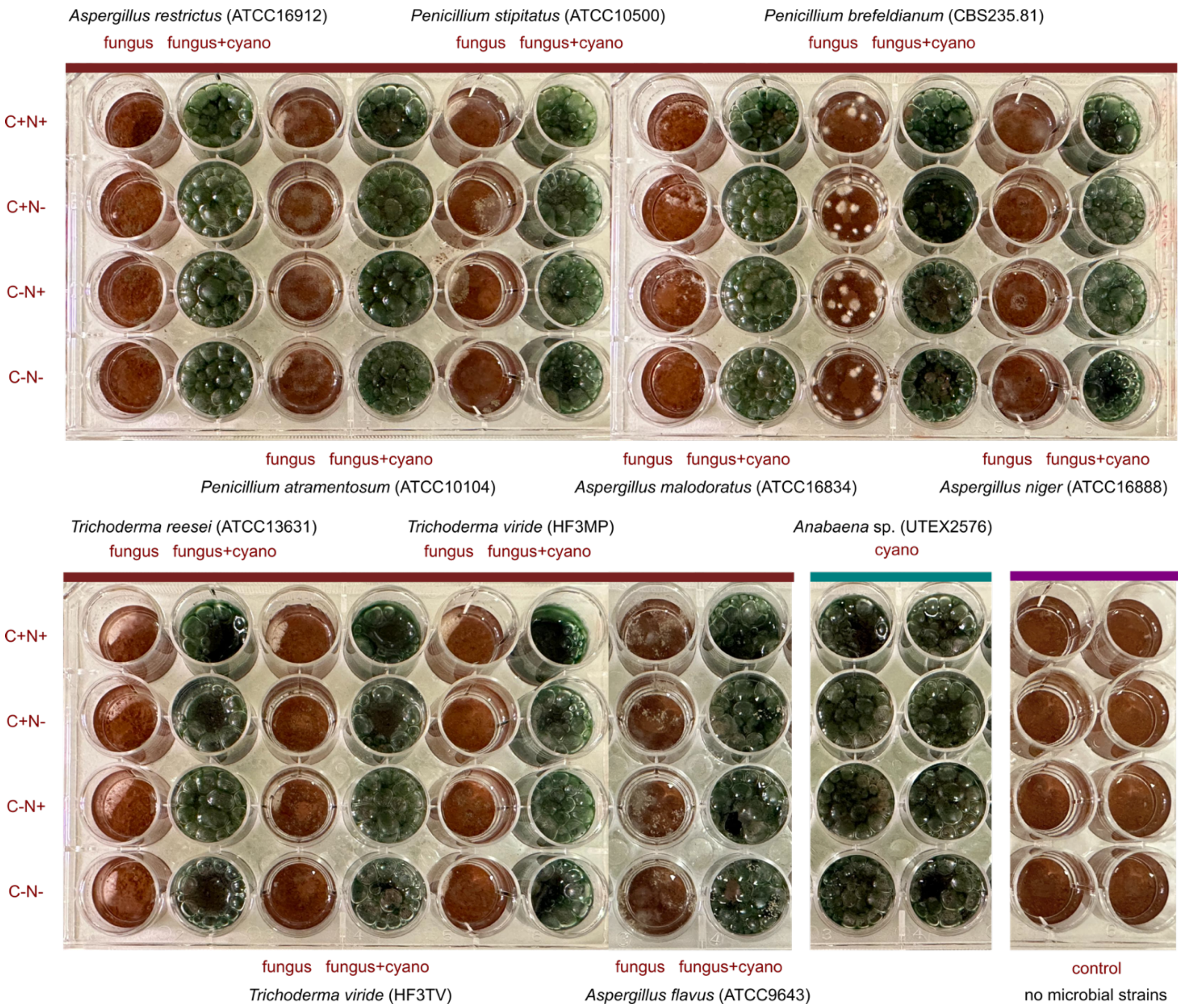

**FIGURE S2.** The macroscopic picture of each well after 28-day incubation.

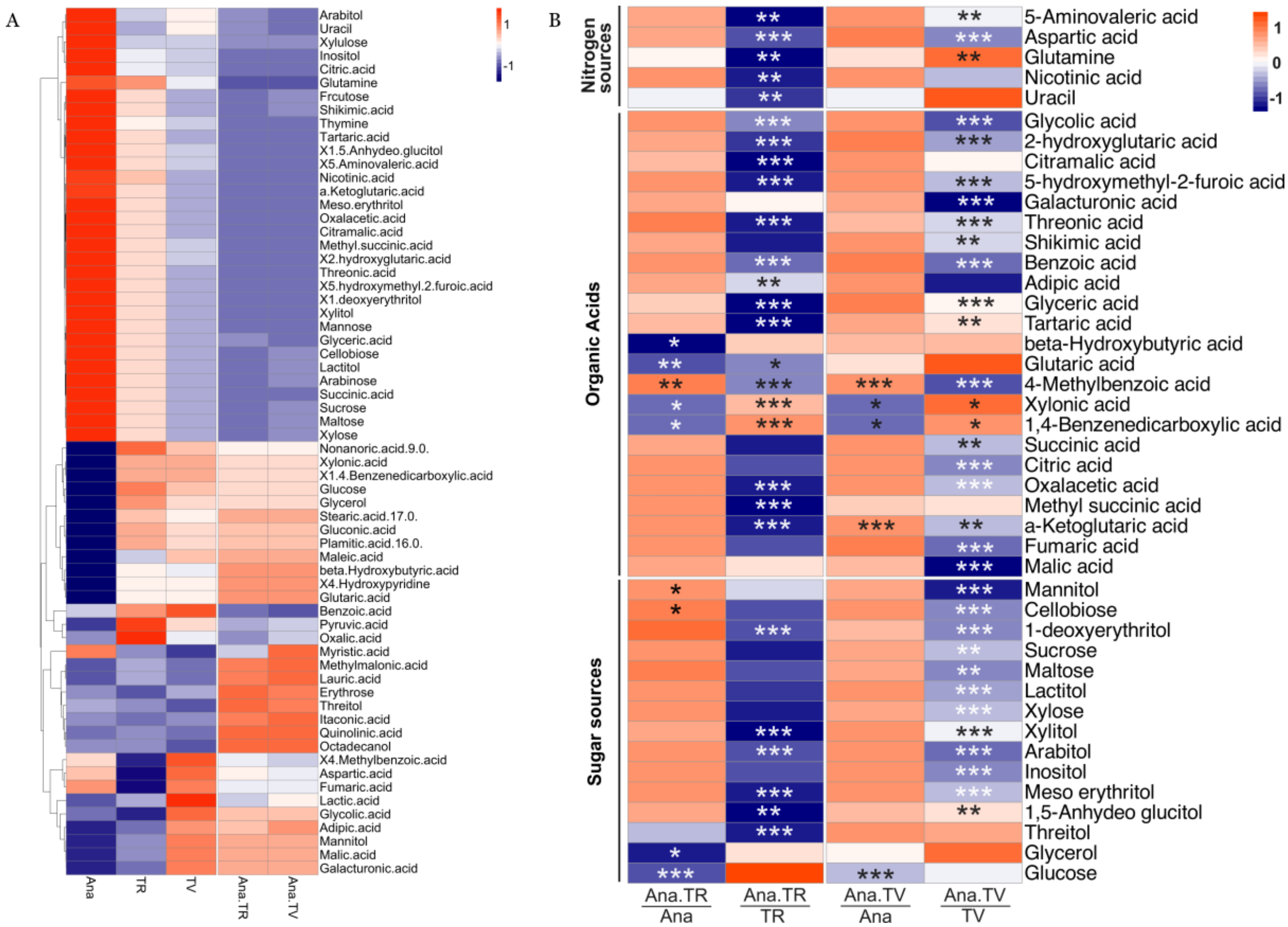

**FIGURE S3.** (A) Heatmap of 64 metabolites selected based on VIP scores derived from PLS-DA analysis, showing Z-score–normalized metabolite abundance across five treatments. (B) Heatmap of differentially expressed metabolites, indicated by asterisks (*$p < 0.05$, **$p < 0.01$, ***$p < 0.001$). Each cell represents the $\log_2$ fold change with Z-score normalization across treatments.